\documentclass[aps,prb,twocolumn,superscriptaddress,showpacs]{revtex4}

\usepackage{graphicx}
\usepackage{color}

\begin{document}

\title{Ambipolar transport in bulk crystals of a topological insulator
by gating with ionic liquid}

\author{Kouji~Segawa}
\email[]{segawa@sanken.osaka-u.ac.jp}

\author{Zhi~Ren}
\author{Satoshi~Sasaki}
\affiliation{Institute of Scientific and Industrial Research, 
Osaka University, Osaka 567-0047, Japan}

\author{Tetsuya~Tsuda}
\affiliation{Frontier Research Base for Global Young Researchers, Graduate School 
of Engineering, Osaka University, 2-1 Yamada-oka, Suita, Osaka 565-0871, Japan.}
\affiliation{Department of Applied Chemistry, Graduate School of Engineering, 
Osaka University, 2-1 Yamada-oka, Suita, Osaka 565-0871, Japan}

\author{Susumu~Kuwabata}
\affiliation{Department of Applied Chemistry, Graduate School of Engineering, 
Osaka University, 2-1 Yamada-oka, Suita, Osaka 565-0871, Japan}
\affiliation{Core Research for Evolutional Science and Technology (CREST), 
Japan Science and Technology Agency (JST), Kawaguchi, Saitama 332-0012, Japan}

\author{Yoichi~Ando}
\email[Corresponding author: ]{y_ando@sanken.osaka-u.ac.jp}
\affiliation{Institute of Scientific and Industrial Research, 
Osaka University, Osaka 567-0047, Japan}

\date{\today}

\begin{abstract}

We report that the ionic-liquid gating of bulk single crystals of a
topological insulator can control the type of the surface carriers and
even results in ambipolar transport. This was made possible by the use
of a highly bulk-insulating BiSbTeSe$_2$ system where the chemical
potential is located close to both the surface Dirac point and the
middle of the bulk band gap. Thanks to the use of ionic liquid, the
control of the surface chemical potential by gating was possible on the
whole surface of a bulk three-dimensional sample, opening new
experimental opportunities for topological insulators. In addition, our
data suggest the existence of a nearly reversible electrochemical
reaction that causes bulk carrier doping into the crystal during the
ionic-liquid gating process.

\end{abstract}

% Not specified yet

% 73.25.+i 	Surface conductivity and carrier phenomena
% 71.18.+y 	Fermi surface: calculations and measurements; effective mass,
%                     g factor
% 73.20.At 	Electron states at surfaces and interfaces - Surface states,
%                 band structure, electron density of states
%
% 72.20.My      Conductivity phenomena in semiconductors and insulators -
%                 Galvanomagnetic and other magnetotransport effects
% 72.20.Ee      Mobility edges; hopping transport
% 74.62.Dh      Effects of crystal defects, doping and substitution
% 73.30.+y      Surface double layers, Schottky barriers, and work functions 

\maketitle

\section{Introduction}

Three-dimensional (3D) topological insulators (TIs) are characterized by
a novel topological order \cite{Moore, Fu-Kane, Roy, QHZ} which dictates
the appearance of spin-filtered massless Dirac fermions on the
surface.\cite{RMP_TI_10, Moore_Nature10, Qi_RMP11} To experimentally
address the peculiar physics of 3D TIs, it is desirable to access the
Dirac point of the surface state (SS).\cite{RMP_TI_10, Moore_Nature10,
Qi_RMP11} This is relatively easy with the surface-sensitive
spectroscopies such as the angle-resolved photoemission
\cite{Hsieh_Nature08, NishidePRB10, Xia_Nphys09, SatoPRL10,
Hiroshima_TBE_PRL10, ChenPRL10, Xue-NatCom, Arakane_NComm12} and the
scanning tunneling microscope,\cite{HanaguriPRB10, PCheng_PRL10} but it
is more challenging for the bulk-sensitive transport experiments because
the chemical potential is always pinned to the bulk bands (including the
impurity band) in real materials. To tune the chemical potential to a
desirable position for transport experiments, two approaches have been
employed: one is the tuning of the chemical compositions upon
synthesizing crystals,\cite{HorPRB09, RenPRB10, Taskin_BSTS_PRL11,
Ren_BSTS_PRB11, Ren_Cd_PRB11, Jia_PRB11} and the other is the gating to
control the surface carriers.\cite{ChenJ_PRL10, Steinberg_NL10,
CheckelskyPRL11, Kong_NNano11, Yuan_NL11} Among the latter approach, the
electric-double-layer gating (EDLG) method is a promising new technique
\cite{Yuan_NL11} to allow application of a large electric
field.\cite{Dhoot}

In the EDLG configuration, either cations or anions in a liquid
electrolyte are accumulated near the surface of a sample by application
of an electric field, and they form an electric double layer which
generates a very strong electric field locally on the surface. This leads
to the induced surface carrier density of as high as $\sim$10$^{15}$
cm$^{-2}$.\cite{Dhoot} Such a large tunability of the surface carrier
density is the merit of the EDLG method. In the context of TIs, Yuan
{\it et al.} showed \cite{Yuan_NL11} that the EDLG method allows an
ambipolar doping control in ultrathin films of Bi$_2$Te$_3$; however,
the surface state was obviously gapped in the ultrathin films used in
Ref. \onlinecite{Yuan_NL11}, and consequently the transport properties
were likely to be dominated by the bulk state. In the present paper, we
show that the EDLG method can achieve ambipolar transport even in bulk
single crystals, if one uses the highly bulk-insulating
Bi$_{2-x}$Sb$_{x}$Te$_{3-y}$Se$_{y}$ (BSTS)
system.\cite{Taskin_BSTS_PRL11, Ren_BSTS_PRB11} In our experiment, the
control of the chemical potential was possible on the {\it whole surface
of a bulk 3D sample}, opening new experimental opportunities for TIs. In
addition, our data suggest that the chemical potential is moving not
only at the surface but also in the bulk, possibly due to a nearly
reversible electrochemical reaction to cause apparent bulk doping during
the EDLG process.

\section{Sample Preparations and measurements}

% Figure 1
\begin{figure*}
\includegraphics*[width=11cm,clip]{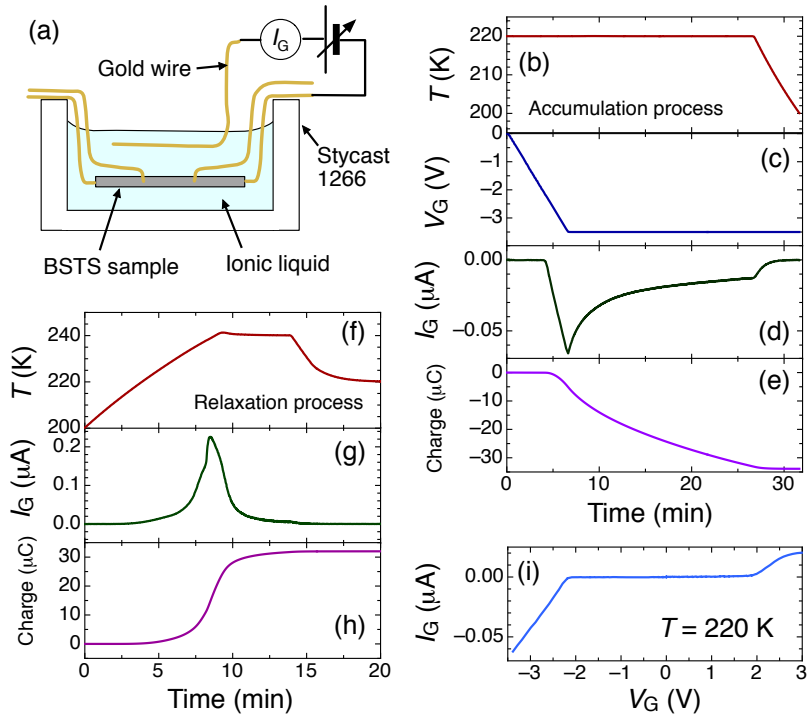}  %8.5cm for double columns
\caption{(color online)
(a) Schematic picture of the experimental setup. Actual ionic liquid has
no color. (b)-(e) Changes in parameters with time in the procedure of
applying the gate voltage $V_G$ with the target value of $-3.5$ V on a
BSTS sample. First, $V_G$ is gradually changed from 0 to $-3.5$ V while
keeping the temperature at 220 K, and we wait for at least 10 min for the
electric double layer to develop; the system is then cooled down, and the
gate current $I_G$ vanishes completely after the ionic liquid
solidifies. (f)-(h) Changes in $T$, $I_G$, and the integrated charge in the
relaxation process from $V_G$ = $-3.5$ V, in which $V_G$ is set to 0 V and
the system is warmed to $\sim$240 K. (i) $V_G$ vs $I_G$ curve at 220 K
with the voltage sweep rate of $\sim$$-0.5$ V/min.}
\end{figure*}

A series of BSTS single crystals were grown by a modified Bridgman
method.\cite{Taskin_BSTS_PRL11, Ren_BSTS_PRB11} Six gold wires were
attached to each sample by a spot welding technique, and magnetotransport
measurements were performed by a conventional ac six-probe method by
sweeping the magnetic field between $\pm$9 T. An electrically insulating
cup made of Stycast 1266 was used as a sample container, in which the
sample with gold wires was submerged into an ionic liquid (IL)
electrolyte, as schematically shown in Fig. 1(a). We used a specially
purified ionic liquid [EtMeIm][BF$_4$] as the electrolyte.\cite{IL} As
the gate electrode, an additional piece of gold wire was dipped into the
IL without touching the sample. In this paper, we adopt the convention
of defining the gate voltage by taking the sample as the reference
point, as was done in most of the previous works. In this convention,
$n$-type carriers are supposed to be doped to the surface when a
positive gate voltage is applied.

The gate voltage was applied in the following procedure: first, the
temperature $T$ of the sample was stabilized at 220 K; then, the gate
voltage $V_G$ was swept slowly, and the temperature was kept at 220 K
for at least 10 min after the voltage reached the set value;
finally, the sample was cooled down slowly. When changing the gate
voltage, we cycled the above procedure. Figures 1(b)-1(e) show an
example of the history of $T$, $V_G$, gate current $I_G$, and the
accumulated charge, during the procedure for the target $V_G$ of $-3.5$
V. With the IL used in the present experiment, gate current was never
detected at temperatures below 200 K, even when the gate voltage was
changed. This is because when the IL solidifies, ions are not mobile at
all. 

The accumulated ions can be released when the gate voltage is set to
zero and the system is warmed up to 220 K. Such a relaxation process
from $V_G$ = $-3.5$ V is shown in Figs. 1(f)-1(h). One can see that the
amount of released charge [$\sim$32 $\mu$C in Fig. 1(h)] is comparable
to that of the accumulated charge [$\sim$34 $\mu$C in Fig. 1(e)]. The
small difference between the released and accumulated charge is probably
an indication of an irreversible electrochemical reaction during the
gating processes at a high gate voltage. In the above example, the total
surface area of the sample was 13.4 mm$^2$, so that the $\sim$32 $\mu$C
of charge accumulated on the surface corresponds to the accumulated ion
density of $\sim$1.6$\times 10^{15}$ cm$^{-2}$ and the capacitance of
the unit area 80 $\mu$F cm$^{-2}$, which is comparable to that
previously reported for EDLG.\cite{YuanZnO} Figure 1(i) shows $V_G$ vs
$I_G$ curve at 220 K, which indicates that there is a threshold $|V_G|$
of $\sim$2 V below which little current flows, suggesting that the ions
are not mobile below this threshold voltage; similar behavior was
previously reported for a different IL.\cite{Ueno} The origin of this
behavior is currently not clear, but it may well be a characteristic of the
ionic liquid used here. Also, it was difficult to obtain reproducible
results of the transport properties for $|V_G|$ between 2 and 3 V,
probably because the formation of the electric double layer is unstable
in this gate-voltage region. Therefore, we closely measured the
transport properties only in the range of $|V_G|\ge 3$ V and $V_G$ = 0
V.

\section{EDLG experiment on $\mathbf{Bi_{1.5}Sb_{0.5}Te_{1.7}Se_{1.3}}$ single crystals}

% Figure 2
\begin{figure}[t]
\includegraphics[width=8.5cm,clip]{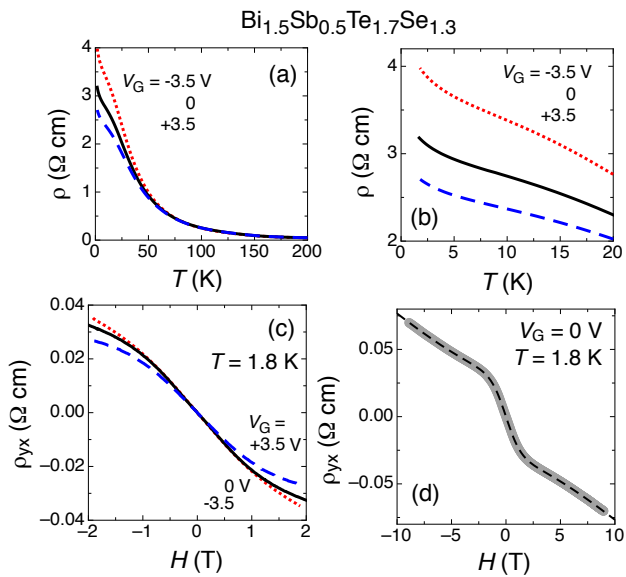}
\caption{(color online)
(a),(b) Temperature dependences of the resistivity of a
Bi$_{1.5}$Sb$_{0.5}$Te$_{1.7}$Se$_{1.3}$ single crystal
for $V_G=-3.5$, 0, and +3.5 V.
(c) Magnetic-filed dependences of $\rho_{yx}$ at 1.8 K.
In these three figures (a)-(c), the dotted, solid, and broken lines correspond to
$V_G$ of $-3.5$, 0, and +3.5 V, respectively.
(d) Fitting of the two-band conduction model to the $\rho_{yx}(H)$ data
at $V_G$ = 0 V, which yields $\rho_{\rm bulk}$ = 3.2 $\Omega \,{\rm
cm}$, $n_{\rm bulk}$ = 9.1 $\times$ 10$^{16}$ cm$^{-3}$, $\mu_{\rm
bulk}$ = 29 cm$^2$/V$\,$s, $\rho_{\rm SS}^{\rm 2D}$ = 8.1 ${\rm
k}\Omega$, $n_{\rm SS}$ = 1.6 $\times$ 10$^{12}$ ${\rm cm}^{-2}$, and
$\mu_{\rm SS}$ = 4.7 $\times$ 10$^{3}$ cm$^2$/V$\,$s.
The gray thick line represents the experimental result and the solid line 
is the fitted result.}
\label{rhochiplot}
\end{figure}

First, we show the results of the EDLG experiments on
Bi$_{1.5}$Sb$_{0.5}$Te$_{1.7}$Se$_{1.3}$, which is currently one of the
most bulk-insulating TI materials. \cite{Taskin_BSTS_PRL11,
Ren_BSTS_PRB11} Figure 2(a) shows the temperature dependences of the
resistivity $\rho$ with the gate voltage ($V_G$) of $-3.5$, 0, and +3.5
V. Even though this is a bulk single crystal, at low temperatures $\rho$
presents a clear change upon application of $V_G$; namely, below
$\sim$50 K, $\rho$ changes with changing $V_G$, and it tends to decrease
(increase) for positive (negative) $V_G$. Figure 2(b) magnifies this
change for $T \le$ 20 K. Figure 2(c) shows the magnetic-field
dependences of the Hall resistivity, $\rho_{yx}(H)$. The sign of the
charge carriers remains negative in this sample, although a significant
non-linearity suggests the coexistence of surface and bulk conduction
channels. Note that the surface conduction in TIs can involve both the
topological surface states and topologically-trivial two-dimensional
electron-gas (2DEG) states that appear as a result of band bending.
\cite{Bianchi} In the present experiment, when the gate
voltage is applied, the 2DEG states are most likely contributing to the
conduction alongside of the topological surface states. Unfortunately,
we have not been able to elucidate the contributions of the topological
and non-topological surface states because no Shubnikov-de Haas (SdH)
oscillation has been observed in our gated samples. (To successfully
separate the contributions of the two, detailed information obtained
from SdH oscillations is necessary.\cite{Taskin_BSTS_PRL11,Taskin_MBE})
We therefore make no claim of the composition of the surface carriers in
the present paper.

In Fig. 2(c), one can see a clear tendency that both the absolute value
of $\rho_{yx}$ and the slope of $\rho_{yx}(H)$ decrease upon increasing
$V_G$ from $-3.5$ to +3.5 V; this means that the apparent electron
concentration increases with increasing $V_G$. Since $n$-type carriers
are expected to be doped when a positive $V_G$ is applied, the present
observation can be understood as a natural consequence of the EDLG. The
nonlinear $H$-dependence observed in $\rho_{yx}$ is useful for gaining
insights into the respective roles of bulk and surface transport
channels, because its analysis based on a simple two-band model
\cite{RenPRB10} gives a crude idea about the relevant transport
parameters. For example, the $\rho_{yx}(H)$ data at $V_G$ = 0 V give the
following estimate based on the fitting shown in Fig. 2(d): 
For the bulk channel, the bulk resistivity
$\rho_{\rm bulk} \simeq$ 3 $\Omega \,{\rm cm}$, the bulk carrier density
$n_{\rm bulk} \simeq$ 9 $\times$ 10$^{16}$ cm$^{-3}$, and the bulk
mobility $\mu_{\rm bulk} \simeq$ 30 cm$^2$/V$\,$s; for the surface
channel, the sheet resistance $\rho_{\rm SS}^{\rm 2D} \simeq$ 8 ${\rm
k}\Omega$, the surface carrier density $n_{\rm SS} \simeq$ 2 $\times$
10$^{12}$ ${\rm cm}^{-2}$, and the surface mobility $\mu_{\rm SS}
\simeq$ 5 $\times$ 10$^{3}$ cm$^2$/V$\,$s. In this fitting, the
constraint imposed by the presence of sharp kinks at $\sim\pm$1 T helps
reduce the ambiguity in the fitting parameters, and in fact, the three
parameters, $n_{\rm bulk}$, $\mu_{\rm bulk}$, and $n_{\rm SS}$, are in
reasonable agreement with our previous transport studies of BSTS
involving SdH oscillations.\cite{Taskin_BSTS_PRL11, Ren_BSTS_PRB11} The
large value of $\mu_{\rm SS}$ would imply that the SdH oscillations be
observed, but we did not observe any SdH oscillations in this sample;
this is possibly because the surface chemical potential (and hence
$n_{\rm SS}$) is not very uniform throughout the sample and the SdH
oscillations with various different frequencies add up to smear visible
oscillations.

The above result of the two-band analysis suggests that the surface
contribution in the total conductance was only $\sim$1\%, which is
reasonable because the measured sample had a considerable thickness of
332 $\mu$m. Nevertheless, if one looks at the EDLG effect in resistivity
[Figs. 2(a) and 2(b)], one notices a very puzzling fact: for the
negative $V_G$ of $-3.5$ V, the number of surface electrons are expected
to be reduced and indeed, the slope of $\rho_{yx}$ gets larger; however,
the resistivity increase is as much as 25\%. Since the surface
contribution in the total conductance is only $\sim$1\%, even when the
surface conduction is completely suppressed by EDLG, one can expect an
increase in $\rho$ of $\sim$1\% at most, as long as the bulk channel is
not affected by EDLG. Therefore, the observed large increase in $\rho$
strongly suggests that the bulk channel must also be affected by EDLG.
Indeed, the $\rho_{yx}(H)$ data shown in Fig. 2(c) presents a clear
change in the slope at high fields for different $V_G$, which implies
that not only $n_{\rm SS}$ but also $n_{\rm bulk}$ is changing. To
corroborate this inference, the two-band analyses of the $\rho_{yx}(H)$
data at finite $V_G$ suggests that the bulk carriers decreases
(increases) by 5\% (10\%) for $V_G$ of $-3.5$ V ($+3.5$ V).

It is very surprising that the EDLG affects the density of bulk carriers
by a noticeable amount in a sample as thick as 332 $\mu$m, but our
transport data can hardly be understood if one does not accept this
possibility. Given that the electric field generated by $\sim$1 $\times$
10$^{15}$ cm$^{-2}$ of ions on the surface is shielded in less than 100
nm,\cite{penetration} the only possibility is that some bulk doping into
the BSTS sample is taking place during the EDLG process. In this
respect, the slow time scale of the change in the measured current
during the EDLG process seems to support the idea that some
electrochemical reaction is taking place. 

In passing, we note that we have not successfully measured SdH
oscillations in any of the gated samples. This is likely to be due to an
inhomogeneous distribution of the local electric field (which is
conceivable because our samples have a lot of macroscopic terraces on
the surface) or some chemical degradation of the surface caused by the
ionic-liquid gating. Hence a more definitive analysis of the transport
data beyond the simple two-band analysis is currently unavailable.
Nevertheless, the bulk doping due to the EDLG seems to be an inevitable
conclusion of our result.

\section{Ambipolar transport in $\mathbf{BiSbTeSe_{2}}$ single crystals}

It was recently found \cite{Arakane_NComm12} that the position of the
chemical potential with respect to the Dirac point of the SS is tunable
within the bulk band gap in BSTS when one follows a particular series of
$x$ and $y$ that were identified in Ref. \onlinecite{Ren_BSTS_PRB11}. In
particular, the chemical potential was found to be close to both the
Dirac point of the SS and the middle of the bulk band gap in BiSbTeSe$_2$.
Therefore, for our EDLG experiment, to maximize the possibility of
achieving ambipolar transport, we mainly measured the BiSbTeSe$_2$
system. (The gating data shown in Fig. 1 were taken during the
experiment on the BiSbTeSe$_2$ sample reported below.)

% Figure 3
\begin{figure}[t]
\includegraphics[width=8.5cm,clip]{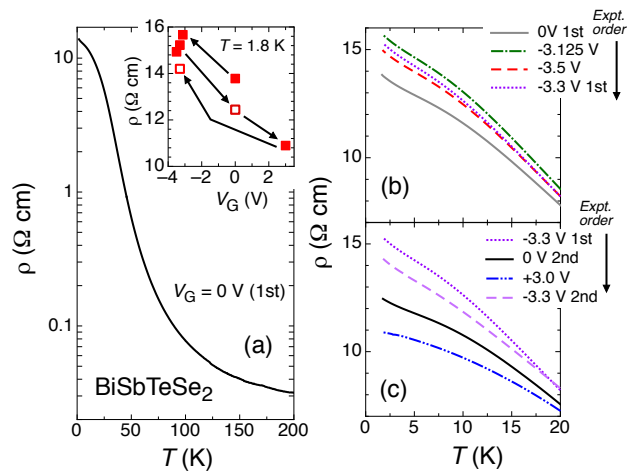} %8.5cm for double columns
\caption{(color online)
(a) Temperature dependence of $\rho$ in the BiSbTeSe$_{2}$ single
crystal used for the EDLG experiment at $V_G$ = 0 V in a semi-log plot;
inset shows the $V_G$ dependence of $\rho$ at 1.8 K, where the arrows indicate
the order of experiments. 
(b),(c) Low-temperature part of $\rho(T)$ for various $V_G$.}
\end{figure}

\begin{table}[]
\centering % used for centering table
\begin{tabular}{c c c c} % centered columns (3 columns)
\hline\hline %inserts double horizontal lines
Experiment  & $V_G$ & $\rho$ ($\Omega$ cm)  & $R_H^{\rm LF}$ (cm$^3$/C) \\ 
order & (V) & at 1.8 K  & at 1.8 K\\ 
[0.5ex] % adds vertical space
\hline % inserts single horizontal line
1 & 0 & 13.8 & $-3940$  \\ % inserting body of the table
2 & $-3.125$ & 15.7 & $-2290$  \\ % inserting body of the table
3 & $-3.5$ & 14.9 & 699  \\ % inserting body of the table
4 & $-3.3$ & 15.2 & 165  \\ % inserting body of the table
5 & 0 & 12.4 & $-1440$  \\ % inserting body of the table
6 & $+3.0$ & 10.9 & $-767$  \\ % inserting body of the table
7 & $-3.3$ & 14.2 & 289  \\ % inserting body of the table
[1ex] % adds vertical space
\hline %inserts single line
\end{tabular}
\caption{Experimental order of applied $V_G$ and resultant transport properties 
at low temperatures.
$R_H^{\rm LF}$ is the low field limit of the Hall coefficient.} % title of Table
\end{table}

Figure 3(a) shows the temperature dependence of the resistivity $\rho$
of a BiSbTeSe$_2$ single crystal before applying the gate voltage. One
can see that $\rho$ exceeds 10 $\Omega$cm at low temperature, testifying
to a high quality of this sample. As was the case with
Bi$_{1.5}$Sb$_{0.5}$Te$_{1.7}$Se$_{1.3}$, at low temperatures we
observed clear change in $\rho$ upon application of $V_G$ [Figs. 3(b) and 3(c)]
even though this is a bulk single crystal. However, it also turned out
that the transport properties do not completely recover after cycling
$V_G$. To illustrate the situation, we show in Table I the measured
transport properties at 1.8 K for various values of $V_G$ in the order
of the measurements. As one can see in this table, the transport
properties for $V_G$ = 0 were measured twice, before and after applying
$V_G$ = $-3.5$ V; also, there are two different data for $V_G$ = $-3.3$
V, which were taken before and after $V_G$ = $+3.0$ V was applied. In
both cases, the resistivity decreased after a high voltage was applied. 
Such a decrease in resistivity after application of a high voltage was
also observed in other samples, so it appears to be an unavoidable
effect in BSTS crystals; this is probably due to some irreversible
electrochemical reaction taking place in the bulk, which gradually 
spoils the bulk-insulating property. However, as one can
see in the inset of Fig. 3(a), a large part of the change in resistivity
in response to $V_G$ is reversible. This reversible part of the change
is consistent with the picture that the number of $n$-type carriers
increases with positive $V_G$ and decreases with negative $V_G$ due to
the EDLG effect.

An interesting feature in our resistivity data is that the resistivity
value presents a maximum around $-3.2$ V [see inset of Fig. 3(a) and also
Fig. 3(b)]; namely, the resistivity is {\it smaller} at $V_G$ of $-3.5$
V compared to that at $-3.3$ V, despite the overall trend that negative
voltage increases $\rho$. In fact, when the system is in the $n$-type
regime, a more negative $V_G$ value would lead to a smaller number of
$n$-type carriers, and one would expect the resistivity to increase; the
opposite behavior observed for $V_G < -3.3$ V suggests that the system
is changing from $n$-type to $p$-type. To confirm this possibility, one
must look at the Hall data.

% Figure 4
\begin{figure}[t]
\includegraphics[width=8.5cm,clip]{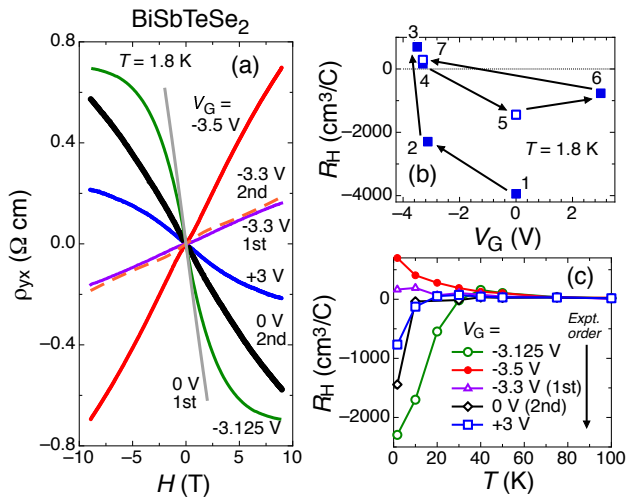} %8.5cm for double columns
\caption{(color online)
(a) Magnetic-field dependences of $\rho_{yx}$ at 1.8 K for various $V_G$.
(b) $V_G$ dependence of $R_H$ at 1.8 K. 
(c) Temperature dependences of $R_H$ for various $V_G$.
Arrows indicate the experimental order.}
\end{figure}

Figure 4(a) shows how the behavior of $\rho_{yx}(H)$ 
changes with $V_G$. Initially, at $V_G$ = 0
V (first) and at $-3.125$ V, the charge carriers are clearly $n$-type,
but the carriers become $p$-type at large negative $V_G$ values of
$-3.5$ and $-3.3$ V. This is a signature of an ambipolar transport in a
bulk crystal, and confirms the idea that the sign change of charge
carriers takes place between $V_G$ = $-3.125$ and $-3.3$ V. As noted
above, the apparent maximum in the resistivity near $V_G \simeq -3.2$ V
is consistent with this interpretation. The $n$-type doping is recovered when
$V_G$ was set to 0 V again, and the slope of $\rho_{yx}(H)$ was found to
decrease with increasing $V_G$ up to $+3.0$ V, suggesting an increase in 
the $n$-type carriers, as expected.

It should be noted that we observed the sign change in this sample again
after setting $V_G$ to $+3.0$ V and then bringing it back to $-3.3$ V,
as shown by a broken line in Fig. 4(a). Thus, the sign change of the
carriers is obviously reproducible. By defining the Hall coefficient
$R_H$ as the slope of $\rho_{yx}(H)$ at low field, we summarize the
gate-voltage dependences of $R_H$ in Fig. 4(b). One can see that the
sign change in $R_H$ is reproducibly observed, although the exact
value of $R_H$ at a given $V_G$ shows a history dependence.

The temperature dependences of $R_H$ for various $V_G$ measured in the
successive five experiments [Fig. 4(c)] indicate that the ambipolar
transport is observed only below $\sim$30 K where thermal activations of
bulk carriers are negligible. Since the experiment on
Bi$_{1.5}$Sb$_{0.5}$Te$_{1.7}$Se$_{1.3}$ discussed in the previous
section indicated that the carrier densities in both the bulk and
surface transport channels are changing with EDLG, it is important to
elucidate whether the sign change of the carriers observed in
BiSbTeSe$_2$ is occurring in the bulk or on the surface, or both. To
infer the origin of the sign change, we have analyzed the $\rho_{yx}(H)$
data for various values of $V_G$ with the simple two-band model.

% Figure 5
\begin{figure}[]
\includegraphics[width=6.5cm,clip]{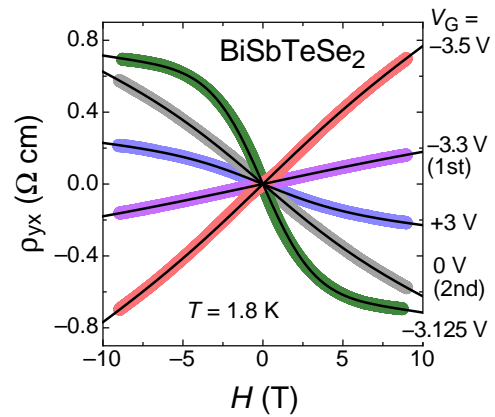} %8.5cm for double columns
\caption{(color online)
Results of the two-band model fitting to the $\rho_{yx}(H)$ data of 
BiSbTeSe$_2$ at various $V_G$. Thick lines are the data at 1.8 K and 
solid lines are the fitting results.}
\end{figure}

The results of the two-band fitting are summarized in Fig. 5 and Table
II. For example, the fitting to the data at $V_G$ = 0 V shown 
in Fig. 5 gives the following crude estimate for the
transport parameters: For the bulk channel, $\rho_{\rm bulk} \simeq$ 13
$\Omega$cm, $n_{\rm bulk} \simeq$ 1 $\times$ $10^{16}$
cm$^{-3}$, and $\mu_{\rm bulk} \simeq$ 40 cm$^2$/V$\,$s; for the surface
channel, $\rho_{\rm SS}^{\rm 2D}$ = $\rho_{\rm SS}/d \simeq$ 30
k$\Omega$, $n_{\rm SS} \simeq$ 2 $\times$ $10^{11}$ cm$^{-2}$, and
$\mu_{\rm SS} \simeq$ 1 $\times$ $10^{3}$ cm$^2$/V$\,$s. Here, $d$ (=
181 $\mu$m) is the thickness of the sample and the sign of charge
carriers is negative for both the bulk and surface channels. Given that
the chemical potential is very close to the Dirac point in
BiSbTeSe$_{2}$, it seems that the estimate of the surface carrier
density indicated in this analysis is reasonable, in spite of the
weakness of the non-linearity in $\rho_{yx}(H)$. With the above
parameters, the contribution of the surface channel to the total
conductance is calculated to be $\sim$2\%.

\begin{table}[]
\centering % used for centering table
\begin{tabular}{c c c c c} % centered columns (3 columns)
\hline\hline %inserts double horizontal lines
$V_G$ & Bulk carrier & $\mu_{\rm bulk}$ & Surface carrier & $\mu_{\rm SS}$ \\ 
(V) & density (cm$^{-3}$) & (cm$^2$/Vs) & density (cm$^{-2}$)  &  (cm$^2$/Vs) \\ 
[0.5ex] % adds vertical space
\hline % inserts single horizontal line
$+3.0$       & $-4.7$ $\times$ 10$^{16}$ & 12 & $-1.5$ $\times$ 10$^{11}$ & 1.2 $\times$10$^{3}$ \\ % inserting body of the table
0          & $-1.3$ $\times$ 10$^{16}$ & 38 & $-2.2$ $\times$ 10$^{11}$ & 1.0 $\times$10$^{3}$ \\ 
$-3.125$ & $-1.8$ $\times$ 10$^{16}$ & 21 & $-2.0$ $\times$ 10$^{11}$ & 2.0 $\times$10$^{3}$ \\ 
$-3.3$      & 5.0 $\times$ 10$^{16}$ & 8 & 1.6 $\times$ 10$^{11}$ & 4.6 $\times$10$^{2}$ \\ 
$-3.5$       & 9.8 $\times$ 10$^{15}$ & 42 & 1.8 $\times$ 10$^{11}$ & 8.4 $\times$10$^{2}$ \\ 
[1ex] % adds vertical space
\hline %inserts single line
\end{tabular}
\caption{Parameters of the two-band model fitting of the $\rho_{yx}(H)$ data
at various $V_G$; the fitted curves are shown in Fig. 5.} 
\end{table}

Looking at the results of the two-band analyses of the data at finite
$V_G$, one can see that the Hall data strongly suggests that both the
bulk and surface carriers change sign simultaneously at $V_G \le -3.3$
V. Indeed, we found that it is impossible to fit the $\rho_{yx}$ data
for $V_G \le -3.3$ V by assuming that only one of the two channels
changes sign. Therefore, it appears that in the present EDLG experiment
the chemical potential is swung from the $n$-type regime to the $p$-type
regime not only on the surface but also in the bulk. Most likely, what
is happening in the bulk at negative $V_G$ is compensation due to
electrochemical $p$-type doping, which eventually overwhelms the
preexisting $n$-type bulk carriers.

% Figure 6
\begin{figure}[t]
\includegraphics[width=6.5cm,clip]{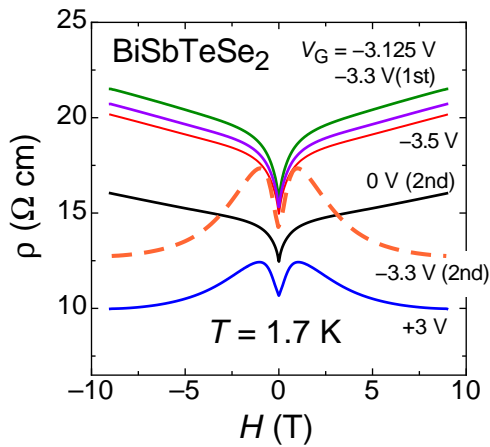} %8.5cm for double columns
\caption{(color online)
Magnetoresistance data at various stages of the EDLG experiment of 
BiSbTeSe$_2$ at 1.7 K.}
\end{figure}

Finally, we show in Fig. 6 the magnetoresistance (MR) data for various
$V_G$ values. Obviously, the data present the weak antilocalization (WAL)
effect at low fields;\cite{WAL1,WAL2,Taskin_MBE} since the WAL effect is
a signature of two-dimensional transport and is not usually observed in
bulk TI crystals dominated three-dimensional transport, the MR data give
additional evidence for a sizable contribution of the surface transport
to the total conductance. At the same time, this WAL effect makes it
difficult to examine the consistency of the parameters obtained from the
two-band analysis of $\rho_{yx}(H)$ in the MR data. Also, the behavior
of MR qualitatively changed after the application of $V_G$ = $+3$ V, the
origin of which is not clear at the moment. Since the MR is too
complicated and not very reproducible, we did not try to make a detailed
analysis.

%\vspace{-0.3cm}

\section{Discussion}

It is important to mention that there are missing charges in our EDLG
experiment; namely, the total amount of charge induced by gating in the
sample is much smaller than the total amount of ions accumulated on the
surface. For example, in the case of BiSbTeSe$_2$, the total amount of
accumulated charges measured by the current is $\sim$34 $\mu$C, which
corresponds to the ion density on the surface of $\sim$1.6 $\times$
10$^{15}$ cm$^{-2}$. On the other hand, the change in the surface
carrier density in BiSbTeSe$_2$ was $\sim$4 $\times$ 10$^{11}$ cm$^{-2}$
and its bulk carrier density changed by less than 1 $\times$ 10$^{17}$
cm$^{-3}$, which amounts to the total charge of less than 20 $\mu$C.
Therefore, obviously the gating is not very efficiently performed. In
this regard, the doping control of the surface carriers in the present
experiment is similar to another EDLG experiment on Bi$_2$Te$_3$ thin
film,\cite{Yuan_NL11} where $\sim$7 $\times$ 10$^{11}$ cm$^{-2}$ of
surface carriers were doped with $V_G \simeq$ $-3$ V. For other
materials, the amount of surface carrier doping by EDLG is of the order
of 10$^{13}$--10$^{15}$ cm$^{-2}$,
\cite{Dhoot,YuanZnO,Ye_PNAS11,EndoAPL10} and thus the electric-field
effect on Bi-based topological insulators appears to be exceptionally
ineffective.\cite{note}

The mechanism to cause these missing charges is not clear at the moment,
but we speculate that some electrochemical redox reaction involving
adsorbed molecules on the surface causes a layer of immobile charges
that shields some fraction of the electric field created by the ions,
leading to a weakening of the electric field for inducing mobile
carriers in the sample.

Also, the bulk doping that accompanies the EDLG in our BSTS samples is
surprising. Remarkably, the data for BiSbTeSe$_2$ suggests that the bulk
doping process is nearly reversible and it takes place in the time scale
of the order of 10 min. The chemical mechanism of this bulk doping is
not clear at the moment, but the possible cause might be intercalation
of ions into the van-der-Waals gap in the BSTS crystal. Obviously, there
is a lot to understand about the electrochemistry accompanying the EDLG
on Bi-based tetradymite TI materials.

\section{conclusion}

In conclusion, the electric-double-layer gating (EDLG) using ionic
liquid was applied to bulk single crystals of BSTS to control the
chemical potential, and ambipolar transport was observed in a sample of
BiSbTeSe$_2$ as thick as 181 $\mu$m. The gating was successfully applied
to tune the chemical potential on the whole surface of a
three-dimensional sample, and surprisingly, it appears that the EDLG on
BSTS crystals is accompanied by a nearly reversible electrochemical
reaction that caused bulk carrier doping. It turned out that the EDLG is
exceptionally inefficient for the BSTS system, with the maximum change
in the surface carrier density of $\sim$4 $\times$ 10$^{11}$ cm$^{-2}$
despite the ion density on the surface of $\sim$1.6 $\times$ 10$^{15}$
cm$^{-2}$. The key to the successful ambipolar carrier control in
the present experiment was the use of BiSbTeSe$_2$ crystals in which
the chemical potential is located close to the middle of the bulk band
gap \cite{Arakane_NComm12} and the residual bulk carrier density was
only $\sim$1 $\times$ 10$^{16}$ cm$^{-3}$. In combination with a
technique to open a gap on the surface,\cite{Nomura,Sato} the present
experiment paves the way for topological magnetoelectric-effect
experiments \cite{QHZ} that require the chemical-potential control on
the whole surface of a bulk topological insulator, although the
mechanism of the bulk doping associated with EDLG needs to be understood
before this technique is comfortably applied.

\section{acknowledgment}

This work was supported by JSPS (NEXT Program), MEXT (Innovative Area
``Topological Quantum Phenomena" KAKENHI 22103004 and KAKENHI 20371297),
and AFOSR (AOARD 104103 and 124038).


\begin{thebibliography}{10}
\parskip-0.2ex plus0.05ex minus0.05ex

\bibitem{Moore}
J.E. Moore and L. Balents, Phys. Rev. B {\bf 75}, 121306(R) (2007).

\bibitem{Fu-Kane}
L. Fu and C.L. Kane, Phys. Rev. B {\bf 76}, 045302 (2007).

\bibitem{Roy}
R. Roy, Phys. Rev. B {\bf 79}, 195322 (2009).

\bibitem{QHZ}
X.-L. Qi, T. L. Hughes, and S.-C. Zhang, Phys. Rev. B {\bf 78}, 195424 (2008).

\bibitem{RMP_TI_10} 
M.Z. Hasan and C.L. Kane, Rev. Mod. Phys. {\bf 82}, 3045 (2010).

\bibitem{Moore_Nature10} 
J.E. Moore, {Nature} {\bf {464}}, 194 ({2010}).

\bibitem{Qi_RMP11} 
X.-L. Qi and S.-C. Zhang, Rev. Mod. Phys. {\bf 83}, 1057 (2011).

\bibitem{Hsieh_Nature08} D. Hsieh, D. Qian, L. Wray, Y. Xia, Y.S. Hor,
R.J. Cava, and M.Z. Hasan, Nature {\bf 452}, 970 (2008).

\bibitem{Xia_Nphys09} Y. Xia, D. Qian, D. Hsieh, L. Wray, A. Pal, H.
Lin, A. Bansil, D. Grauer, Y.S. Hor, R.J. Cava, and M.Z. Hasan, Nature
Physics {\bf 5}, 398 (2009).

\bibitem{NishidePRB10} A. Nishide, A.A. Taskin, Y. Takeichi, T. Okuda,
A. Kakizaki, T. Hirahara, K. Nakatsuji, F. Komori, Y. Ando, and I.
Matsuda, Phys.\ Rev.\ B {\bf 81}, 041309(R) (2010).

\bibitem{SatoPRL10} T. Sato, K. Segawa, H. Guo, K. Sugawara, S. Souma,
T. Takahashi, and Y. Ando, Phys.\ Rev.\ Lett. {\bf 105}, 136802 (2010).

\bibitem{Hiroshima_TBE_PRL10} K. Kuroda, M. Ye, A. Kimura, S.V. Eremeev,
E.E. Krasovskii, E.V. Chulkov, Y. Ueda, K. Miyamoto, T. Okuda, K.
Shimada, H. Namatame, and M. Taniguchi, Phys.\ Rev.\ Lett. {\bf 105},
146801 (2010).

\bibitem{ChenPRL10} Y.L. Chen, Z.K. Liu, J.G. Analytis, J.-H. Chu, H.J.
Zhang, B.H. Yan, S.-K. Mo, R.G. Moore, D.H. Lu, I.R. Fisher, S.C. Zhang,
Z. Hussain, and Z.-X. Shen, Phys.\ Rev.\ Lett. {\bf 105}, 266401 (2010).

\bibitem{Xue-NatCom} J. Zhang, C.-Z. Chang, Z. Zhang, J. Wen, X. Feng,
K. Li, M. Liu, K. He, L. Wang, X. Chen, Q.-K. Xue, X. Ma, and Y. Wang,
Nat. Commun. {\bf 2}, 574 (2011). 

\bibitem{Arakane_NComm12} T. Arakane, T. Sato, S. Souma, K. Kosaka, K.
Nakayama, M. Komatsu, T. Takahashi, Z. Ren, K. Segawa, and Y. Ando, Nat.
Commun. {\bf 3}, 636 (2012).

\bibitem{HanaguriPRB10} 
T. Hanaguri, K. Igarashi, M. Kawamura, H. Takagi, and T. Sasagawa, 
Phys.\ Rev.\ B {\bf 82}, 081305(R) (2010).

\bibitem{PCheng_PRL10} P. Cheng, C. Song, T. Zhang, Y. Zhang, Y. Wang,
J.-F. Jia, J. Wang, Y. Wang, B.-F. Zhu, X. Chen, X. Ma, K. He, L. Wang,
X. Dai, Z. Fang, X. Xie, X.-L. Qi, C.-X. Liu, S.-C. Zhang, and Q.-K. Xue,
Phys.\ Rev.\ Lett. {\bf {105}}, 076801 ({2010}).

\bibitem{HorPRB09} Y.S. Hor, A. Richardella, P. Roushan, Y. Xia, J.G.
Checkelsky, A. Yazdani, M.Z. Hasan, N.P. Ong, and R.J. Cava, Phys.\ Rev.\
B {\bf 79}, 195208 (2009).

\bibitem{RenPRB10} Z. Ren, A.A. Taskin, S. Sasaki, K. Segawa, and Y.
Ando, Phys. Rev. B {\bf 82}, 241306 (2010).

\bibitem{Taskin_BSTS_PRL11} A.A. Taskin, Z. Ren, S. Sasaki, K. Segawa,
and Y. Ando, Phys. Rev. Lett. {\bf 107}, 016801 (2011).

\bibitem{Ren_BSTS_PRB11} Z. Ren, A.A. Taskin, S. Sasaki, K. Segawa, and
Y. Ando, Phys. Rev. B {\bf 84}, 165311 (2011).

\bibitem{Ren_Cd_PRB11} Z. Ren, A.A. Taskin, S. Sasaki, K. Segawa, and Y.
Ando, Phys. Rev. B {\bf 84}, 075316 (2011).

\bibitem{Jia_PRB11} S. Jia, H. Ji, E. Climent-Pascual, M.K. Fuccillo,
M.E. Charles, J. Xiong, N.P. Ong, and R.J. Cava, Phys. Rev. B {\bf {84}}, 235206
({2011}).

\bibitem{Steinberg_NL10} H. Steinberg, D.R. Gardner, Y.S. Lee, and P.
Jarillo-Herrero, Nano Lett. {\bf 10}, 5032 (2010).

\bibitem{ChenJ_PRL10} J. Chen, H.J. Qin, F. Yang, J. Liu, T. Guan, F.M.
Qu, G.H. Zhang, J.R. Shi, X.C. Xie, C.L. Yang, K.H. Wu, Y.Q. Li, and L.
Lu, Phys. Rev. Lett. {\bf 105}, 176602 (2010).

\bibitem{CheckelskyPRL11} J.G. Checkelsky, Y.S. Hor, R.J. Cava, and N.P.
Ong, Phys. Rev. Lett. {\bf 106}, 196801 (2011).

\bibitem{Kong_NNano11} D. Kong, Y. Chen, J.J. Cha, Q. Zhang, J.G.
Analytis, K. Lai, Z. Liu, S.S. Hong, K.J. Koski, S.-K. Mo, Z. Hussain,
I.R. Fisher, Z.-X. Shen, and Y. Cui, Nat. Nanotechnol. {\bf {6}},
705 (2011).

\bibitem{Yuan_NL11} H. Yuan, H. Liu, H. Shimotani, H. Guo, M. Chen, Q.
Xue, and Y. Iwasa, Nano Lett. {\bf {11}}, 2601 (2011).

\bibitem{Dhoot} A. S. Dhoot, J. D. Yuen, M. Heeney, I. McCulloch, D. Moses
and A. J. Heeger, Proc. Natl. Acad. Sci. {\bf 103}, 11834 (2006); M. J.
Panzer, and C. D. Frisbie, Adv. Func. Mat. {\bf 16}, 1051 (2006).

\bibitem{IL} T. Tsuda, K. Kondo, T.
Tomioka, Y. Takahashi, H. Matsumoto, S. Kuwabata, and C. L. Hussey,
Angew. Chem. Int. Ed. {\bf 50}, 1310 (2011).

\bibitem{YuanZnO}
H. Yuan, H. Shimotani, A. Tsukazaki, A. Ohtomo, M. Kawasaki and Y. Iwasa,
Adv. Funct. Mater. {\bf 19}, 1046 (2009).

\bibitem{Ueno}
K. Ueno, S. Nakamura, H. Shimotani, H. T. Yuan, N. Kimura, T. Nojima,
H. Aoki, Y. Iwasa, and M. Kawasaki, Nat. Nanotechnol. {\bf 6}, 408 (2011).

\bibitem{Bianchi}
M. Bianchi, D. Guan, S. Bao, J. Mi, B. Brummerstedt Iversen, 
P. D. C. King, P. Hofmann, Nat. Commun. {\bf 1}, 128 (2010).

\bibitem{Taskin_MBE} 
A.A. Taskin, S. Sasaki, K. Segawa, and Y. Ando, 
Phys. Rev. Lett. {\bf 109}, 066803 (2012).

\bibitem{penetration}
K. Ueno, S. Nakamura, H. Shimotani, A. Ohtomo, N. Kimura, T. Nojima, H. Aoki, 
Y. Iwasa and M. Kawasaki,
Nature Materials {\bf 7}, 855 (2008).

\bibitem{Ye_PNAS11}
J. Ye, M.F. Craciun, M. Koshino, S. Russo, S. Inoue, H. Yuan, H. Shimotani, 
A.F. Morpurgo and Y. Iwasa,
Proc. Natl. Acad. Sci. U.S.A. {\bf 108}, 13005 (2011).

\bibitem{EndoAPL10}
M. Endo, D. Chiba, H. Shimotani, F. Matsukura, Y. Iwasa and H. Ohno,
Appl. Phys. Lett. {\bf 96}, 022515 (2010).

\bibitem{note}
There is a report that $\sim$1.3 $\times$ 10$^{14}$ cm$^{-2}$ of
carriers were doped to Bi$_2$Se$_3$ films by EDLG [Y. Onose {\it et al.}, 
Appl. Phys. Express {\bf 4}, 083001 (2011)], but the 6-nm films
studied in that work were heavily Se deficient and contained as
much as 2.2 $\times$ 10$^{20}$ cm$^{-3}$ of electrons.
Hence, the efficiency of the EDLG on TI samples may be improved in more
metallic samples.

\bibitem{WAL1}
J. Chen, X. Y. He, K. H. Wu, Z. Q. Ji, L. Lu, J. R. Shi, J. H. Smet, and Y. Q. Li, 
Phys. Rev. B {\bf 83}, 241304(R) (2011).

\bibitem{WAL2}
H. Steinberg, J. B. Lal\"{o}e, V. Fatemi, J. S. Moodera, and P. Jarillo-Herrero, 
Phys. Rev. B {\bf 84}, 233101 (2011).

\bibitem{Nomura}
K. Nomura and N. Nagaosa, Phys. Rev. Lett. {\bf 106}, 166802 (2011).

\bibitem{Sato}
T. Sato, K. Segawa, K. Kosaka, S. Souma, K. Nakayama, K. Eto, T. Minami, 
Y. Ando, and T. Takahashi, Nature Phys. {\bf 7}, 840 (2011).


\end{thebibliography}
\end{document}